# A Hybrid Image Encryption Scheme based on Chaos and a DPA-Resistant Sbox

Mohammad Gholamzadeh, Behrouz Khadem

## Abstract

Image encryption is one of the most common and effective methods to secure digital images. Recently, Khalid M. Hosny presented an image encryption scheme based on 6D hyper chaotic mapping and Q-Fibonacci matrix, which, despite its remarkable theoretical and practical properties, has several weaknesses, including inaccuracy of black image encryption, inappropriate white image encryption (improper entropy parameters, correlation, chi-square test, histogram, UACI, and NPCR), weak keys, inappropriate key usage. In this paper, based on Khaled Hosny's design, a new effective design is presented that has improved encryption security and efficiency. In addition, in the proposed design, a secure key and a substitution box with a high degree of transparency order, which is resistant to DPA attacks, have been added. Also, a method to improve transferring chaos parameters is also proposed. The test results show the improvement of the resistance of the proposed design against the common attacks of image encryption schemes and improvement in bandwidth consumption. Also it has been shown that the proposed scheme has produced better results in terms of both security and efficiency compared to other similar new schemes.

# سامانه ترکیبی رمز تصویر مبتنی بر آشوب و جعبه جانشانی مقاوم در DPA

محمد غلامزاده [1]، بهروز خادم [2]

[1] کارشناس ارشد مخابرات امن و رمزنگاری، دانشگاه جامع امام حسین (ع)
fr6ad@ihu.ac.ir

[2] دانشیار دانشکده کامپیوتر، شبکه و ارتباطات، دانشگاه جامع امام حسین (ع)
bkhadem@ihu.ac.ir

## چکیده

رمز تصویر از متداول‌ترین و مؤثرترین روش‌های امن‌سازی تصاویر دیجیتال است. اخیراً خالد حسنی یک طرح رمز تصویر مبتنی‌بر نگاشت ابر آشوب ۶ بعدی و ماتریس Q-فیبوناچی ارائه کرده است که علی‌رغم خواص نظری و عملی قابل توجه، به دلیل کم‌توجهی به عامل آشفته‌سازی، قابلیت رمزنگاری تصویر کاملاً سیاه و کاملاً سفید یا هر تصویر کاملاً یکنواخت دیگر را ندارد. استفاده نادرست از پارامترهای اولیه نگاشت آشوبی و وجود کلیدهای ضعیف و ضعف در کاربرد کلید از دیگر نقاط ضعف طرح خالد حسنی است. در این مقاله بر مبنای طرح خالد حسنی، یک طرح جدید ارائه شده که نگاشت ابر آشوب و ماتریس Q-فیبوناچی را بهبود داده است. به‌علاوه در طرح جدید یک کلید و یک جعبه‌جانشانی با مرتبه شفافیت بالا که در برابر حملات DPA مقاوم است، افزوده شده است. همچنین روشی برای بهبود ارسال پارامترهای آشوب نیز پیشنهاد شده است. نتایج حاصل از آزمایش‌ها، نشان‌دهنده ارتقای مقاومت طرح جدید در مقابل حملات متداول رمزهای تصویر و کاهش مصرف پهنای باند است.

## کلمات کلیدی

رمز تصویر، آشوب، نگاشت آشوبی، جعبه‌جانشانی امن، مرتبه شفافیت

## ۱- مقدمه

در سال‌های اخیر، کاربرد گسترده انتقال تصاویر از طریق شبکه‌های مجازی باعث افزایش توجه به رمزنگاری تصویر شده است. به‌منظور محرمانگی تصویر و حفظ حریم خصوصی کاربران مطالعات عمیق و گسترده‌ای روی طرح‌ها و سامانه‌های رمزنگاری که خواص انتشار و آشفتگی قابل‌قبول داشته باشند به‌عمل‌آمده است.

آشوب، یکی از انواع دینامیک‌های غیرخطی است که در طبیعت (مانند آب و هوا[1])، دینامیک ماهواره‌ها در منظومه شمسی و رشد جمعیت در بوم‌شناسی) و در آزمایشگاه (مانند مدارهای الکتریکی، لیزرها، واکنش‌های شیمیایی، دینامیک‌های شار و سیستم‌های مکانیکی و مغناطیسی) قابل‌مشاهده است. پدیده‌های آشوبی در کاربردهای متعدد دیگری نظیر هواشناسی، فیزیک، مهندسی رایانه، رمزنگاری، مهندسی مخابرات و الکترونیک، فناوری‌های ارتباطی و اطلاعاتی، پزشکی و زیست‌شناسی هم قابل‌مشاهده است [2-5]. آشوب برای اولین بار توسط متیوز در رمزنگاری ادغام شد [6, 7] و به نظر می‌رسد فردریک در [8] برای اولین بار پیشنهاد داد تا از یک نگاشت آشوبی برای رمزنگاری تصویر استفاده شود.

در دهه‌های اخیر آشوب و رمزنگاری تصویر توجه گسترده ای را به خود جلب کرده است [2, 5, 9-13]. برخی از خواص شناخته‌شده و اصلی سامانه‌های آشوبی وابستگی خروجی آن‌ها به شرایط اولیه، شبه‌تصادفی بودن، توپولوژیک بودن [1] و چگال بودن [2] خروجی آن‌ها است که همگی از الزامات اساسی هر سیستم رمز کلاسیک شانونی است. بنابراین می توان از آنها برای طراحی سیستم‌های رمزنگاری استفاده کرد [5, 12, 14, 15].

یک دنباله آشوبی هم می‌تواند در تولید جعبه‌ی جانشانی [16, 17] و هم در طراحی سامانه رمزنگاری تصویر [18, 19] برای ایجاد اغتشاش و

انتشار به‌کار برود. بنابراین، رمزنگاری تصویر با استفاده از جعبه‌های جانشانی و سیستم آشوب امکان پذیر است [20, 21].

در ریاضیات، دنباله فیبوناچی دنباله‌ای از اعداد است که در آن اعداد (غیر از دو عدد اول) با محاسبه مجموع دو عدد قبلی ایجاد می‌شوند[22]. ماتریس Q-فیبوناچی[3] به صورت ماتریس $Q = \begin{pmatrix} 1 & 1 \\ 1 & 0 \end{pmatrix}$ است، رابطه(1) توان nام ماتریس Q-فیبوناچی را نشان می‌دهد(ویژگی‌ای وجود دارد که ماتریس Q را به اعداد فیبوناچی متصل می کند).

$$Q^n = \begin{pmatrix} F_{n+1} & F_n \\ F_n & F_{n-1} \end{pmatrix} \quad (1)$$

اخیراً در[12]، یک طرح رمزنگاری تصویر با استفاده از نگاشت ابر آشوب 6 بعدی و ماتریس Q-فیبوناچی توسط خالد حسنی[4] و همکاران پیشنهاد شده است. دراین طرح (که از این پس طرح IEAHF نامیده می‌شود) مقادیر اولیه (کلیدها) نگاشت آشوبی، قبل از شروع به‌کار سامانه رمزنگاری تولید و بین فرستنده و گیرنده توافق می‌شوند. در IEAHF دنباله آشوبی حاصل از تکرار یک سامانه دینامیکی 6 بعدی، پس از مرتب سازی وارد مرحله انتشار (توسط ماتریس Q-فیبوناچی) می‌شود و در نهایت تصویر رمزشده خروجی پس از (حداقل) دو بار تکرار در حلقه درونی سامانه تولید می‌شود.

در ادامه این مقاله به شرح زیر سازمان‌دهی شده است در بخش اول اصطلاحات و مفاهیم بنیادی تشریح شده است. بخش دوم شامل مروری بر IEAHF است. در بخش سوم، به ارزیابی نقاط ضعف IEAHF پرداخته شده است. در بخش چهارم طرح پیشنهادی GH401 نیز مطرح می‌شود. شبیه‌سازی نرم‌افزاری GH401 و مقایسه با سایر طرح‌های مشابه، برتری امنیت و کارایی طرح پیشنهادی را نشان می‌دهد. در بخش انتهایی مقاله است نتیجه‌گیری کلی ارائه می‌شود.

## 2- اصطلاحات و مفاهیم بنیادی

مهم‌ترین معیارهای امنیت در یک سامانه رمز تصویر مانند هیستوگرام، حملات تفاضلی شامل NPCR و UACI[5]، توزیع مربع کای پیرسون[7]، آنتروپی و همبستگی تصویر رمزی هستند[23].

### 2-1- تحلیل هیستوگرام

یک الگوریتم رمزنگاری تصویر خوب باید به‌گونه‌ای باشد که در آن هیستوگرام متن رمزی تقریباً یکنواخت است و به تصویر تصادفی نزدیک‌تر باشد [24]. برای جلوگیری از نشت اطلاعات و جلوگیری از حملات مهاجم، مهم است که تضمین شود تصویر اصلی و تصویر رمز هیچ‌گونه تشابه آماری ندارند.

### 2-2- معیارهای امنیتی شامل NPCR و UACI

در این حمله، هدف مهاجم رمزگشایی تصاویر رمزی (بدون استفاده از کلید) از طریق تعیین رابطه بین تصویر اصلی با تصویر رمزی متناظرش است. معمولاً برای تعیین میزان رابطه بین پیکسل‌های تصویر اصلی ورودی بر روی پیکسل‌های تصویر رمزی خروجی از معیارهای NPCR و UACI استفاده می‌شود [25].

### 2-3- تحلیل توزیع مربع کای

یکنواختی توزیع تصویر رمزی را می‌توان به‌صورت کمی با استفاده از آزمون مربع کای ارزیابی کرد. در این آزمون توزیع مربع کای، جهت مقایسه زیبندگی[8]فراوانی‌های مشاهده‌شده در یک نمونه اندازه‌گیری شده با فراوانی‌های مورد انتظار در توزیع مقایسه می‌شود.

آماره‌ی این آزمون عبارت است از:

$$x^2_{test} = \sum_{k=1}^{256} \left( \frac{v_k - e_k}{e_k} \right)^2 \quad (2)$$

که $k$ تعداد حالت‌های ممکن یک پیکسل و $v_k$ فراوانی هر سطح خاکستری (0-255) است[26]. به طور مثال، اگر تصویر اصلی دارای ابعاد H×W باشد آنگاه $e_k = \frac{H \times W}{256}$ با فرض میزان بحرانی 0/01، اگر $x^2_{test} > x^2(255, 0.01)$، آنگاه نتیجه گرفته می‌شود که فرضیه یکنواختی رد می‌شود و روشن می‌شود که توزیع هیستوگرام تصویر رمز غیر یکنواخت است.

### 2-4- تحلیل آنتروپی

آنتروپی، میزان تصادفی بودن خروجی یک سیستم رمزنگاری را معلوم می‌کند، هر چه آنتروپی اطلاعات یک تصویر رمزی بیشتر باشد، توزیع اطلاعات تصادفی‌تر است و اطلاعات بصری آن کمتر است. در حالت تصادفی ایده‌آل، هرکدام از این پیکسل‌ها در تصویر رمزی باید دارای احتمال وقوع یکسانی باشند. رفتار تصادفی تصویر رمزی با آنتروپی مشخص می‌شود. هنگامی‌که پیکسل‌ها (8-بیتی) رمز می‌شوند، آنتروپی در شرایط ایده‌آل باید 8 باشد [12]، اگر یک الگوریتم رمز سمبل‌هایی با آنتروپی اطلاعات کمتر از 8 تولید کند، احتمال قابل‌پیش‌بینی بودن وجود دارد که برای امنیت رمزنگاری تصویر یک تهدید محسوب می‌شود. نحوه محاسبه مقدار در رابطه زیر است:

$$H(m) = \sum_{i=1}^{2^w - 1} P(m_i) \log_2 \frac{1}{p(m_i)} \quad (3)$$

### 2-5- همبستگی

آزمون همبستگی، مهم‌ترین فن مورداستفاده در به‌دست‌آوردن شباهت بین دو تصویر است. احتمال آنکه اطلاعات تصویر اصلی توسط مهاجم کشف شود، با افزایش میزان همبستگی در تصویر رمزی، افزایش می‌یابد. هرچه میزان همبستگی پیکسل‌های رمزی به یکدیگر به صفر نزدیک‌تر باشد، مناسب‌تر است. این معیار به سه نوع همبستگی افقی، همبستگی عمودی و همبستگی قطری دسته‌بندی می‌شود که در تصویر اصلی و تصویر رمز شده هرکدام به‌صورت جداگانه محاسبه می‌شود. جهت محاسبه ضریب همبستگی از رابطه زیر استفاده می‌شود. در این روابط، x و y در معادله نشان‌دهنده دو مقدار

پیکسل متوالی است و N نشان‌دهنده تعداد جفت پیکسل انتخاب شده است [۲۶].

$$E(x) = \frac{1}{N}\sum_{i=1}^{N} x_i \qquad (4)$$

$$D(x) = \frac{1}{N}\sum_{i=1}^{N} \left(x_i - E(x)\right)^2 \qquad (5)$$

$$\text{cov}(x,y) = \frac{1}{N}\left(\sum_{i=1}^{N}\left(x_i - E(x)\right)\left(y_i - E(y)\right)\right) \qquad (6)$$

$$r_{xy} = \frac{\text{cov}(x,y)}{\sqrt{D(x)D(y)}} \qquad (7)$$

## ۳- مروری بر طرح IEAHF

در این طرح مطابق الگوریتم(۱) یک تصویر (اصلی) به‌اندازه M×N پیکسل، مورد رمزنگاری قرار می‌گیرد.

در گام‌های ۱ تا ۳ مقادیر اولیه (کلید) سامانه آشوبی معرفی شده است.

در گام ۴ باتوجه‌به پارامترهای معرفی شده، شش دنباله‌ی آشوبی به طول M×N حاصل می‌شود که دنباله‌های فرد آن به‌صورت سریال (پشت سر هم) الحاق می‌شود و برای گام ۷ بکار می‌روند (اگر دنباله آشوبی ۶ بعدی را به‌صورت ماتریسی دارای ۶ ستون و L ردیف در نظر بگیریم ستون‌های فرد انتخاب شده است.) از دنباله‌های زوج در IEAHF استفاده‌ای نشده است.

یکی از ضعف‌های الگوریتم(۱) عملیاتی است که در فاصله بین گام ۵ و ۶ انجام می‌گردد. ضعف این عملیات در قسمت ۴-۳ (ارزیابی کلیدهای طرح) به طور کامل تشریح شده است.

در گام ۵ با مرتب‌کردن دنباله‌های فرد به‌صورت صعودی، اندیس موقعیت آن‌ها (مکان پیکسل‌ها) در یک بردار جدید بنام S و سپس در یک ماتریس بنام SS با M×N سطر و r ستون ذخیره می‌شود(r تعداد تکرار حلقه درونی الگوریتم است).

در گام‌های ۶ و ۷ و ۸ دنباله آشوبی به‌صورت زیر بلوک‌های ۲×۲ بازسازی شده و در ماتریس Q-فیبوناچی ضرب می‌شود. یکی از نقاط ضعف این الگوریتم تاثیر نداشتن ماتریس Q-فیبوناچی در دور اول می‌باشد که در قسمت ۵-۲-۲ به اصلاح آن پرداخته شده است

در گام‌های ۹ و ۱۰ تصویر رمزی حاصل(خروجی تکرار اولین حلقه) به تکرارحلقه بعدی فرستاده می‌شود تا مجددا گام‌های ۱ تا ۱۰ تکرار شود و تصویر رمز شده نهایی بدست آید.

باتوجه‌به اینکه در هیچ‌یک از گام‌های الگوریتم(۱) آشفتگی لازم برای رمز کردن یک تصویر یکنواخت (سفید یا سیاه) به وجود نمی‌آید، لازم است یک عامل آشفتگی به این الگوریتم اضافه شود که می‌تواند یک جعبه‌جانشانی باشد.

اضافه‌کردن چنین عامل آشفتگی باید به‌گونه‌ای انجام شود که تهدید جدیدی به سامانه رمز اضافه نکند، به این معنی که در مقابل حملاتی مانند حمله DPA و سایر حملات مقاوم باشد.

## ۴- ارزیابی طرح IEAHF

در این قسمت، طرح IEAHF را از سه دیدگاه اصلی بررسی می‌نماییم. دیدگاه اول ارزیابی کلی است که در آن نشان می‌دهیم تصاویر سیاه، سفید و تصاویر یکنواخت به‌درستی رمز نمی‌شوند و برای اصلاح آن مطابق شکل(۶) اقداماتی را انجام می‌دهیم. در دیدگاه دوم به ارزیابی کلیدهای IEAHF می‌پردازیم و نشان می‌دهیم به دلیل استفاده نادرست از کلیدها، محدودیت‌هایی در رمزنگاری به وجود می‌آید و پیشنهادی برای اصلاح آن ارائه می‌شود. در دیدگاه سوم به موضوع روش نادرست انتقال کلید و مصرف بیش از اندازه پهنای باند می‌پردازیم و برای اصلاح آن راهکاری پیشنهاد می‌کنیم.

**الگوریتم (۱): الگوریتم رمزنگاری طرح IEAHF [۱۲]**

| | |
|---|---|
| 1 : | i=1 |
| 2 : | Transform the image array to a vector P. |
| 3 : | Calculate the initial key of the hyperchaotic system as follows: $x_1 = \frac{\sum_{i=1}^{MN} P(i)+(M\times N)}{2^{23}+(M\times N)}$, $x_i = \text{mod}(x_{i-1}\times 10^6, 1)\ i=2,3,...,6$ With the initial conditions; $x_1, x_2, ..., x_6$. |
| 4 : | Iterate the hyperchaotic system in(1) $N_0 + MN/3$ times then discard the $N_0$ values to make a new sequence L with size $M\times N$. (we select three sequences ($x_1, x_3, and\ x_5$) from the system in (1)). |
| 5 : | Sort L in ascending order and return their positions in vector S. |
| 6 : | Permit the image vector P to generate newly shuffled sequence R as follows: $R_i = P(S_i), i=1:MN$ |
| 7 : | Convert the sequence R into the matrix $R'$ and divide it into sub-blocks, each with size $2\times 2$. |
| 8 : | Get the Chipper image C by multiplying each $2\times 2$ sub-block in $R'$ ، در ماتریس فیبوناچی مرتبه ۱۰ ($Q^{10}$): $\begin{pmatrix} C_{i,j} & C_{i,j+1} \\ C_{i+1,j} & C_{i+1,j+1} \end{pmatrix} = \begin{pmatrix} R'_{i,j} & R'_{i,j+1} \\ R'_{i+1,j} & R'_{i+1,j+1} \end{pmatrix}\begin{pmatrix} 89 & 55 \\ 55 & 34 \end{pmatrix} \mod 256$ $with\ i=1:3:......:M, j=1:3:......:N.$ |
| 9 : | Let I=C then $i = i+1$. |
| 10 : | Replicates steps 2 TO 8 for i<= 2. |

## 4-1- ارزیابی کلی

عدم صحت رمزنگاری تصویر سیاه یکی از ضعف‌های IEAHF (الگوریتم ۱) است. در شکل (۱) تصویر اصلی سیاه به‌عنوان ورودی به این طرح داده شده است. تصویر ورودی به الگوریتم بدون هیچ تغییری در خروجی ظاهر می‌شود که نشان‌دهنده ضعف طرح در بکارگیری مولفه‌های اغتشاش و آشفتگی است.

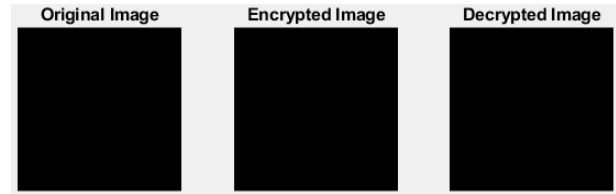

**شکل (۱): عدم رمزنگاری تصویر سیاه در تمام دورها**

به‌علاوه در شکل (۲) نشان‌داده‌شده است؛ تصویر سفید در تکرار اول حلقه الگوریتم(۱) به‌درستی رمز نمی‌شود. این ضعف در دورهای بالاتر نیز خود را نشان می‌دهد که در جداول (۱) تا (۳) قابل‌مشاهده است.

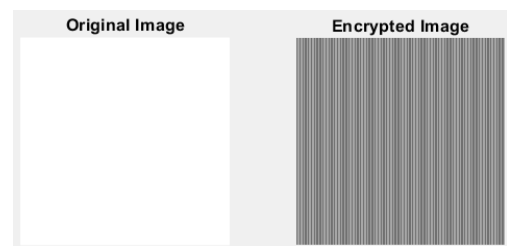

**شکل (۲): تصویر سفید اصلی و رمز شده آن در دور اول الگوریتم(۱)**

با مقایسه هیستوگرام تصویر سفید رمز شده در شکل(۳) و توزیع مربع خی(کای) دو در جدول (۱) عدم یکنواختی در تصویر رمزی طرح IEAHF امکان حملات آماری را به مهاجم می‌دهد.

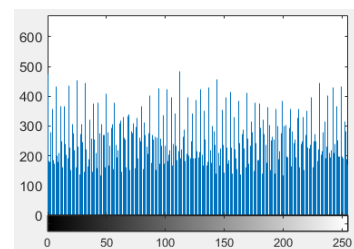

**شکل ۳: عدم یکنواختی هیستوگرام تصویر سفید رمز شده در دور ۵ طرحIEAHF**

مقدار آنتروپی الگوریتم(۱) نمایش‌داده‌شده در جدول (۱) نشان می‌دهد که مقدار پیکسل‌ها در تصویر رمزی سفید، دارای احتمال وقوع یکسانی نیست. ازآنجایی که میزان تصادفی بودن در تصویر سفید رمز شده نامناسب است، احتمال قابل‌پیش‌بینی بودن تصویر رمزی وجود دارد که برای امنیت رمزنگاری تصویر یک تهدید محسوب می‌شود.

**جدول (۱): نامناسب‌بودن مقدار آنتروپی و مربع خی دو به ازای تعداد دورهای مختلف الگوریتم(۱)**

در جدول (۲) مقادیر UACI و NPCR طرح IEAH نشان‌داده‌شده است. در قسمت ۱-۲ ذکر شد یکی از شرایط امنیت، مقاوم بودن در برابر حملات تفاضلی UACI و NPCR است که باتوجه‌به نامناسب‌بودن این

| تصویر سفید ۲۵۶×۲۵۶ | دورها | X2 | Entropy |
|---|---|---|---|
| تصویر اصلی | | ۱۶۷۱۱۶۸۰ | ۰ |
| تصویر رمزی | ۱ | ۸۳۲۳۰۷۲ | ۱ |
| | ۲ | ۲۵۶۵۹۱۵ | ۲٫۷۴۷۵ |
| | ۳ | ۵۹۱۳۶۱٫۷ | ۴٫۹۳۱۵ |
| | ۴ | ۱۱۱۴۶۲٫۳ | ۷٫۰۰۸۰ |
| | ۵ | ۶۷۸۶٫۵ | ۷٫۹۲۷۹ |

شاخص‌ها، معلوم می‌شود که طرح IEAHF در برابر حملات تفاضلی مقاوم نیست.

**جدول (۲): نامناسب‌بودن NPCR، UACI به ازای تعداد دورهای مختلف الگوریتم(۱)**

هرچه میزان همبستگی(سطری، ستونی و قطری) تصویر رمزی بیشتر باشد، اطلاعاتی که مهاجم کشف می‌کند نیز بیشتر است. جدول (۳) همبستگی تصویر سفید رمز شده در الگوریتم (۱) را نشان می‌دهد.

| تصویر سفید ۲۵۶×۲۵۶ | دورها | UACI | NPCR |
|---|---|---|---|
| تصویر رمزی | ۱ | ۰٫۰۰۱۷٪ | ۰٫۰۰۳۰٪ |
| | ۲ | ۴۱٫۵۶۰۴٪ | ۷۴٫۶۹۷۸٪ |
| | ۳ | ۳۲٫۴۰۹۳٪ | ۹۵٫۲۷۷۰٪ |
| | ۴ | ۳۳٫۵۳۲۸٪ | ۹۸٫۸۶۶۲٪ |
| | ۵ | ۳۳٫۵۵۰۰٪ | ۹۹٫۵۵۴۴٪ |

**جدول (۳): مقایسه همبستگی افقی، عمودی و قطری به ازای تعداد دورهای مختلف طرح IEAHF**

به‌طورکلی طرح IEAHF برای رمزنگاری تصاویر یکنواخت مناسب نیست. به‌منظور اثبات این مطلب مطابق شکل (۴) تصویری ۲۵۶×۲۵۶ با

| تصویر سفید ۲۵۶×۲۵۶ | دورها | همبستگی | | |
|---|---|---|---|---|
| | | D | V | H |
| تصویر رمزی | ۱ | ۱ | ۱ | ۱ |
| | ۲ | ۰٫۰۲۱۴۲ | ۰٫۰۲۴۶۵ | ۰٫۵۰۱۹۶ |
| | ۳ | ۰٫۰۱۰۲۳ | ۰٫۰۲۸۷۰ | ۰٫۰۳۳۱۷ |
| | ۴ | ۰٫۰۰۶۱۰ | ۰٫۰۰۶۴۳ | ۰٫۰۳۳۰۸ |
| | ۵ | ۰٫۰۱۵۴۰ | ۰٫۰۰۰۷۳ | ۰٫۰۰۰۵۳ |

پیکسل‌های ۱۲۸ را به الگوریتم (۱) دادیم، نتایج جدول (۱) حاکی از آن است که به ازای هیچ دوری تصویر رمزشده خروجی دارای امنیت نیست.

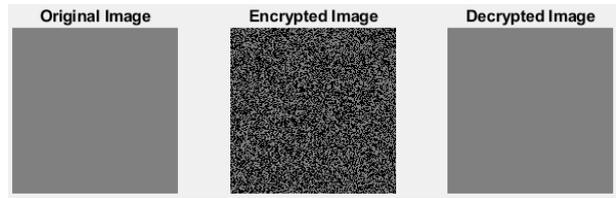

**شکل (۴): تصویر اصلی۲۵۶×۲۵۶ با پیکسل‌های ۱۲۸ و رمز شده آن در دور ۳۰**

ازآنجایی‌که مطابق جدول(۴) مشاهده می‌شود که مقدار آنتروپی تصویر رمز شده بسیار نامناسب است. به همین دلیل احتمال قابل‌پیش‌بینی بودن وجود دارد که برای امنیت رمزنگاری تصویر یک تهدید محسوب می‌شود.

**جدول (۴): نامناسب‌بودن NPCR، UACI و آنتروپی تصویر با مقدار پیکسل‌های ۱۲۸ به ازای تعداد دورهای مختلف طرح IEAHF**

| Entropy | UACI | NPCR | دور | تصویر ۲۵۶×۲۵۶ با پیکسل‌های ۱۲۸ |
|---|---|---|---|---|
| ۰.۹۹۹۹۹۴ | ۳۵.۸۱۷۲٪ | ۹۴.۰۹۶۳٪ | ۲۰ | تصویر رمزی |
| ۰.۹۹۹۹۸ | ۳۷.۳۰۵۶٪ | ۹۸.۱۹۶۴٪ | ۳۰ | |

به‌علاوه هیستوگرام تصویر رمزشده در شکل(۵) نیز عدم امنیت و کارایی طرح IEAHF را نشان می‌دهد.

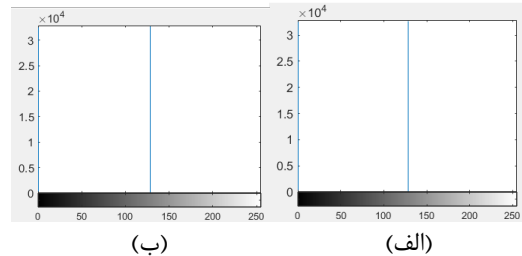

**شکل (۵): هیستوگرام تصویر رمز شده با مقدار پیکسل‌های یکنواخت ۱۲۸ الف) در دور ۲۰ ب) در دور ۳۰**

## ۴-۲- ارزیابی کلیدها

باتوجه‌به محدودیت ارقام قابل نمایش در روش ممیز شناور اعداد حقیقی در نرم‌افزار رایانه و حساسیت شدید نگاشت‌های آشوبی به مقدار ورودی نگاشت، دنباله‌های تولیدشده توسط این نگاشت‌ها در رایانه، پس از یک دوره زمانی معین به‌تدریج خواص آشوبی خود را از دست می‌دهند [۲۷]. به همین دلیل برای جلوگیری از این مشکل یا کاهش تأثیر آن، لازم است در زمان به‌کارگیری این نگاشت‌ها تمهیدات خاصی به کار برود. که در ادامه راهکاری برای رفع این مشکل ارائه می‌شود.

به‌علاوه در طرح[۱۲] (IEAHF) یکی از پارامترهای نگاشت آشوبی ۶ بعدی ($N_0$) به‌عنوان کلید انتخاب شده است و چنانچه در انتخاب مقادیر این کلید توجه کافی نشود، موجب اخلال در صحت عملکرد طرح رمز می‌شود (انجام نشدن گام چهارم الگوریتم۱). به همین دلیل در طرح پیشنهادی عملاً پارامتر $N_0$ از مجموعه کلیدها خارج شده است.

## ۳-۴- انتقال نادرست کلید و مصرف بیش از حد پهنای باند

همان‌طور که در قسمت ۳ گفته شد، یکی از ضعف‌های الگوریتم(۱) مربوط به عملیاتی است که در فاصله بین گام ۵ و ۶ انجام می‌گردد[۲۸] که در [۱۲] توضیح روشنی درمورد آن داده نشده است.

مطابق الگوریتم(۱)در گام ۴ با انتخاب مرتب دنباله‌های فرد (اگر ۶ دنباله را به‌صورت ماتریسی دارای ۶ ستون و L ردیف در نظر بگیریم ستون‌های فرد انتخاب شده است.) و در گام ۵ با مرتب‌کردن آن‌ها به‌صورت صعودی، موقعیت آن‌ها (مکان پیکسل‌ها) در یک بردار جدید بنام S ذخیره می‌شود. سپس طی عملیاتی در[۲۸] باتوجه‌به حلقه درونی در SS (SX) ذخیره شده است.

پس از رمزنگاری تصویر رمزی و ارسال آن، آرایه کلید (که قبلاً از کانال امن ارسال شده) در طرف گیرنده موجب عمل رمزگشایی می‌شود. مشکل [۲۸] آنجایی است که فرستنده به‌جای ارسال کلید مخفی در کانال امن، SS (SX) را روی کانال ارسال کرده تا عمل رمزگشایی سریع‌تر انجام شود. این رویه به شدت مصرف پهنای باند کانال امن را افزایش می‌دهد.

ابعاد آرایه SX بسته به تعداد تکرار حلقه تغییر می‌کند. به طور مثال اگر تعداد تکرار n=4 باشد؛ این آرایه دارای ۴ بعد است که در هر بعد آن دنباله‌ای به طول M×N قرار دارد. برای تصویر اصلی۲۵۶×۲۵۶ با n=4، ۳۳۳ برابر نسبت به حالتی که مقادیر کلید را در کانال امن انتقال دهیم، پهنای باند را بیشتر اشغال می‌نماید(با ابعاد تصویر ۲۵۶×۲۵۶ و n=20، ۸۳۲۱,۵ برابر پهنای باند را بیشتر اشغال می‌نماید). هر چه n بزرگ‌تر باشد پهنای باند بیشتری اشغال می‌شود.

## ۵- طرح پیشنهادی GH401

در این بخش باتوجه‌به نتایج ارزیابی طرح IEAHF و تعیین نقاط ضعف آن که در بخش قبل انجام شد، طرح جدیدی بنام GH401 پیشنهاد شده است که ابتدا ساختار کلی آن معرفی می‌شود، سپس به ارزیابی طرح پیشنهادی با سایر طرح‌های مشابه پرداخته می‌شود و در نهایت نتایج ارزیابی و مقایسه ارائه می‌شود.

## ۱-۵- ساختار کلی

همان‌طور که در شکل(۶) نشان‌داده‌شده است، در ساختار طرح GH401 با اصلاح نگاشت ابر(مافوق) آشوب ۶ بعدی و ماتریس Q-فیبوناچی، ابتدا IEAHF بهبودیافته را معرفی می‌کنیم. سپس در ادامه با طراحی کلید جدید در الگوریتم(۱)، لازم است برای افزایش آشفتگی از یک جعبه جانشانی مانند AES استفاده کنیم، اما باتوجه‌به آسیب‌پذیری آن در برابر حمله DPA، از جعبه جانشانی JAYAR [۲۹] که در برابر حمله DPA مقاوم است،

استفاده می‌کنیم. در بخش‌های بعد با مقایسه نتایج حاصل از آزمایشات طرح GH401 با طرح IEAHF و چندین طرح مشابه دیگر، بهبود اجرا شده و برتری در کارایی و امنیت نیز نشان‌داده‌شده است.

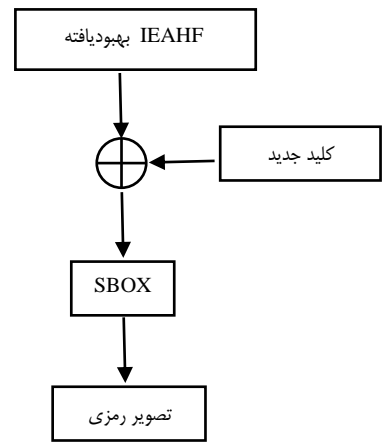

شکل (۶): ساختار طرح پیشنهادی(GH401)

## ۵-۲- اصلاحات اعمال شده

به‌منظور برطرف‌کردن نقاط ضعف IEAHF بهبودهایی به شرح زیر انجام شده و نتیجه اصلاحات در شکل (۸) آورده شده است.

### ۵-۲-۱- اصلاح نحوه استفاده از ابر آشوب ۶ بعدی

یکی از علت‌های رمز نشدن تصویر سیاه و ضعف در رمزنگاری تصویر سفید(به‌طورکلی تصاویر یکنواخت)، استفاده نادرست از نگاشت آشوبی است. به‌منظور رفع این ضعف، رابطه‌ای بین گام ۶ در الگوریتم(۱) و شماره تکرارهای حلقه (n) برقرار می‌کنیم.

$$R_i = P(s_i) + round, i = 1:M \times N, round = 1:n$$

### ۵-۲-۲- اصلاح در روش استفاده از ماتریس Q-فیبوناچی

ازآنجایی‌که ماتریس Q-فیبوناچی در دور اول رمزنگاری طرح IEAHF اثری ندارد؛ به‌منظور بهبود الگوریتم(۱) در گام ۷، ابتدا دنباله R را به ماتریس $R'$ به ابعاد M×N تبدیل می‌نماییم. سپس برای آنکه از همان تکرار اول، ماتریس Q-فیبوناچی در رمزنگاری موثر باشد؛ عدد ثابت ۱ را با $R'$ جمع می‌نماییم و آن را $R''$ می‌نامیم. با تقسیم ماتریس $R''$ به زیربلوک‌های ۲ × ۲ و اضافه کردن عدد ثابت ۱ به هر درایه آن، ماتریس Q-فیبوناچی از تکرار اول حلقه وارد فرایند رمزنگاری می‌شود (نتیجه در شکل (۷) برای تصویر سیاه قابل‌مشاهده است.)

$$\begin{pmatrix} R''_{i,j} & R''_{i,j+1} \\ R''_{i+1,j} & R''_{i+1,j+1} \end{pmatrix} = \begin{pmatrix} R'_{i,j} & R'_{i,j+1} \\ R'_{i+1,j} & R'_{i+1,j+1} \end{pmatrix} + 1$$

$$A = \begin{pmatrix} 89 & 55 \\ 55 & 34 \end{pmatrix}$$

$$fz = Cx \times A$$

$$\begin{cases} C(i,j) = fz(1,1)+1 \\ C(i,j+1) = fz(1,2)+1 \\ C(i+1,j) = fz(2,1)+1 \\ C(i+1,j+1) = fz(2,2)+1 \end{cases} \quad i,j = 1:2:M$$

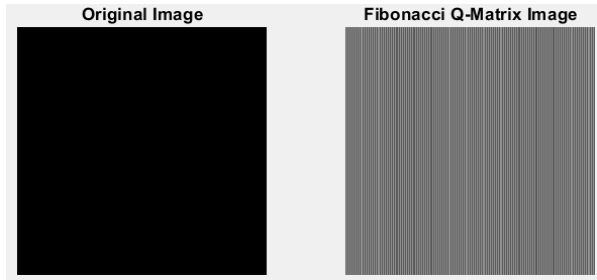

شکل (۷): تصویر اصلی و رمزی پس از اعمال ماتریس Q-فیبوناچی در دور اول

### ۵-۲-۳- افزودن جعبه‌جانشانی مقاوم در برابر حمله DPA

افزودن یک جعبه‌جانشانی به‌عنوان یک مؤلفه اغتشاش دهنده به‌عنوان یک نوآوری در طرح پیشنهادی GH401 انجام شده است. برای جلوگیری از حمله DPA، جعبه جانشانی به طور خاص از [۲۹] انتخاب شده است.(شکل۶).

### ۵-۳- افزودن کلید جدید

به‌منظور صرفه‌جویی در زمان اجرا در ساختار کلید جدید از دنباله‌های زوج ابر آشوب ۶ بعد در گام چهارم الگوریتم(۱) استفاده شده است، (اگر ۶ دنباله تولیدی را به‌صورت ماتریسی دارای ۶ ستون و L ردیف در نظر بگیریم ستون‌های زوج برای کلید انتخاب شده است). به این ترتیب فضای این کلید $2^{16\times 8} = 2^{128}$ بیت است.

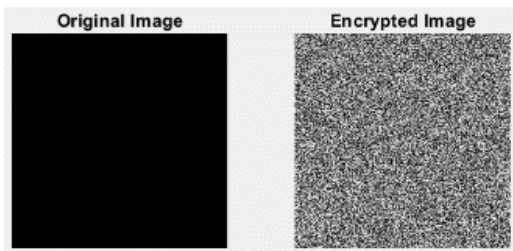

(الف)

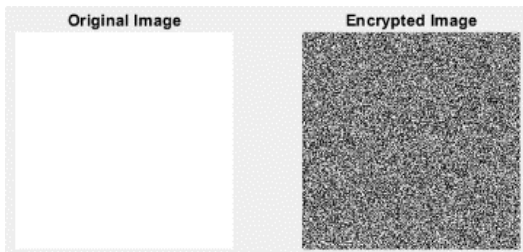

(ب)

شکل (۸): تصویر اصلی و رمز شده طرح GH401 (الف) سیاه در دور چهارم (ب) سفید در دور چهارم

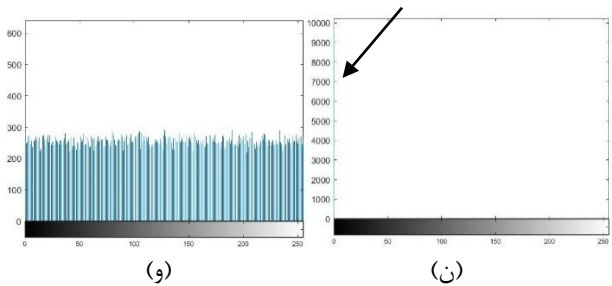

شکل (۹) : الف) عدم یکنواختی هیستوگرام تصویر سفید طرح IEAHF در دور چهارم (ب) عدم یکنواختی هیستوگرام تصویر سفید طرح IEAHF در دور پنجم، (ج) یکنواختی هیستوگرام تصویر سفید طرح GH401 در دور سوم ، (د) یکنواختی هیستوگرام تصویر سفید طرح GH401در دور چهارم، (ن) عدم یکنواختی هیستوگرام تصویر سیاه طرح IEAHF در دور چهارم، و) یکنواختی هیستوگرام تصویر سیاه طرح GH401 در دور چهارم

در شکل (۱۰) دیده می‌شود که توزیع هیستوگرام تصویر رمزی لنا از طرح[۳۰] به وضوح یکنواخت نیست، این در حالی است که توزیع هیستوگرام تصویر رمزی لنا از طرح GH401 کاملا یکنواخت است.

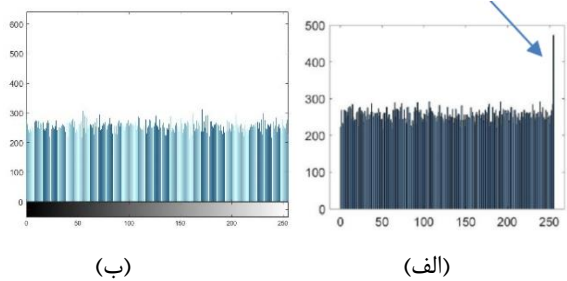

شکل (۱۰): الف) عدم یکنواختی هیستوگرام تصویر لنا پس از رمزنگاری طرح[۳۰] ب) یکنواختی هیستوگرام تصویر لنا پس از رمزنگاری طرح GH401

### ۵-۴-۲- حملات تفاضلی شامل NPCR و UACI

بـرای بررسی ایـن معیار در طـرح IEAHF ، دو تصـویر اصلی از تصـور سفید $PI_1, PI_2$ را در نظر می‌گیریم که تنها در یک پیکسل اختلاف دارنـد. تصـویرهای رمـزی بـه دسـت آمـده و متنـاظر بـا ایـن دو را بـه ترتیب $CI_1, CI_2$ می‌نامیم. همان طور که از جدول(۲) مشاهده می‌شود، در طـرح IEAHF مقـدار حساسیت دو تصویر رمـزی $CI_1, CI_2$ (حتـی در تکرارهای دوم به بعد هم) کم است. این در حالی است که در جـدول (۵) در طرح GH401، دو تصویر رمزی به طـور کامـل از هـم متمایز هسـتند کـه نشان‌دهنده مقاومت طرح GH401 در برابر حملات تفاضلی است.
بـرای تصاویر ۲۵۶×۲۵۶ حداقل NPCR باید ۹۹٫۵۶۹۳ و UACI باید بین (۳۳٫۶۴۴۷% , ۳۳٫۲۸۲۴%) قـرار داشـته باشـد. بهتـرین نتیجـه بـرای UACI، ۳۳٫۴۶۳۵% است[۷, ۱۲].

اضافه‌شدن کلید جدید ضمن افزایش امنیت فضای کلید در برابر حملـه جستجوی فراگیر، حساسیت بالاتری نسبت به کلید مخفی را به وجود می‌آورد و نقص طـرح IEAHF مطرح شـده در ۴-۲ را نیز رفع می‌کنـد. پـس از اصلاحات فوق، تصاویر سیاه، سفید و تصاویر یکنواخت (شکل ۸)، بدرستی رمز می‌شوند.

### ۵-۳-۱- اصلاح کلیدها

علاوه بر طراحی کلید جدید و حذف ($N_r$ به دلیل بیان شـده در قسـمت ۴-۲) در طرح پیشنهادی به‌منظور تکمیل شدن اصلاحات کلید، عـدد شـماره حلقه دوری n به مجموعه کلیدها اضافه شده است. در انتها بـه‌منظور رفع مشکل پهنـای بانـد، بجـای انتقـال SS در کانـال امـن، پارامترهـای $x_1, x_2, x_3, x_4, x_5, x_6, a, b, c, d, e, r, n$ به‌همراه کلید جدید تولید شده از دنباله آشوبی ۶ بعدی را (در کانال امن) برای گیرنده انتقال می‌دهیم.

## ۵-۴- نتایج ارزیابی و مقایسه

تعدادی از مهم‌ترین معیارهای امنیت در یک طرح رمز تصویر مانند حساسیت تصویر رمزی نسبت به تغییرات تصویر اصلی، آنتروپی، هیستوگرام تصویر، همبستگی و توزیع مربع کای پیرسون هستند[۲۳] که در قسمت ۲ هر یـک تشریح شده اند. در ادامه به ارزیابی و مقایسه این معیارها بین طرحIEAHF و GH401 می‌پردازیم ضمنا به‌منظور نشان‌دادن برتری طـرح پیشـنهادی بـا سایر طرح‌های مشابه نیز ارزیابی و مقایسه ای انجام می‌شود.

### ۵-۴-۱- هیستوگرام

همان‌طور که در شکل(۹) مشـاهده می‌شـود عـدم یکنـواختی هیستوگرام در تصویر رمزی سفید و تصویر سیاه طرح IEAHF امکان حملات آماری را به مهاجم می‌دهد. این در حالی است کـه در طـرح GH401 تصویر اصلی و رمزی هیچ‌گونه تشابه آماری ندارند و هیستوگرام تصویر کاملاً یکنواخت است.

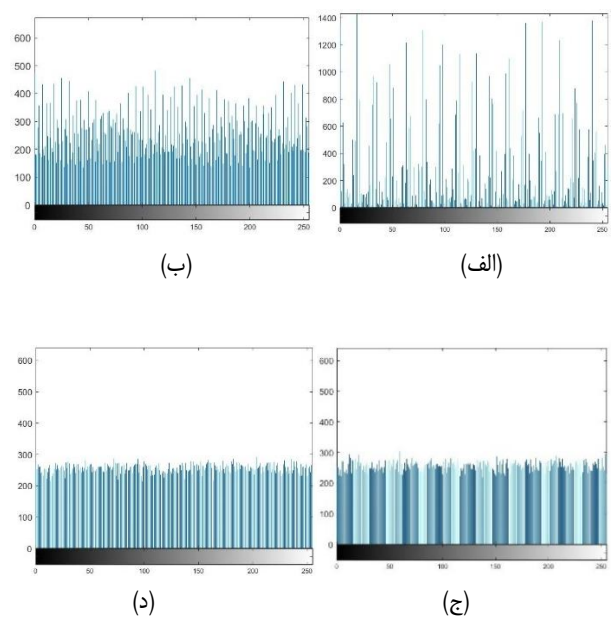

جدول (5): مقادیر NPCR، UACI به ازای تعداد دورهای مختلف طرح GH401

### 4-4-5- آنتروپی

باتوجه‌به نتایج حاصل از طرح IEAHF که در جدول(1) مشاهده می‌شود؛ در طرح Gh401 میانگین آنتروپی 10 تصویر رمزی تصادفی در جدول (6) آمده است که نشان‌دهنده تصادفی بودن طرح است.
مقادیر آنتروپی اطلاعات محاسبه شده تصاویر رمزگذاری شده توسط طرح GH401 در (جدول 11) نشان‌داده‌شده است. مقادیر آنتروپی اطلاعات تصاویر رمزگذاری شده توسط الگوریتم GH401 ما به وضوح نزدیک به 8 و بهتر از سایر الگوریتم‌های رمزنگاری دیگر است.

### 5-4-5- همبستگی

با توجه جدول(3) طرح IEAHF در چند دور ابتدایی تصویر رمزی دارای اطلاعاتی از تصویر اصلی است که احتمال آنکه اطلاعات واقعی توسط مهاجم کشف شود؛ با افزایش میزان همبستگی، افزایش‌یافته است. این درحالی است که در جدول (7) نشان‌داده‌شده است، همبستگی طرح GH401 نزدیک به صفر است.

| تصویر 256×256 | دورها | میانگین UACI | میانگین NPCR% | بهترین NPCR% |
|---|---|---|---|---|
| تصویر رمزی سیاه | 3 | 32٫8501 | 97٫2508 | 97٫5204 |
|  | 4 | 33٫4597 | 99٫6093 | 99٫6483 |
| تصویر رمزی سفید | 3 | 32٫9359 | 97٫8297 | 97٫9415 |
|  | 4 | 33٫4769 | 99٫6145 | 99٫6590 |

برای ارزیابی طرح GH401 از نظر امنیت و کارایی، نتایج با طرح‌های دیگر مقایسه می‌شوند(جدول 10). نتایج تجربی بر اساس میانگین 100 تکرار نشان می‌دهد که GH401 به تصویر اصلی بسیار حساس است و دارای بالاترین نرخ عبور در مقایسه با سایر طرح‌های مشابه است.

### 3-4-5- توزیع مربع کای

باتوجه به آزمون مربع کای، عدم یکنواختی رمز شده تصویر سفید در طرح IEAHF در جدول(1) و یکنواختی طرح GH401 در جدول(6) قابل‌مشاهده است.

جدول (6): مناسب بودن Entropy و مربع کای به ازای تعداد دورهای مختلف طرح GH401

### جدول (7): مقایسه همبستگی افقی، عمودی و قطری به ازای تعداد دورهای مختلف طرح GH401

| تصویر 256×256 | دورها | همبستگی | | |
|---|---|---|---|---|
| | | $D_{Avg}$ / $D_{best}$ | $V_{Avg}$ / $V_{best}$ | $H_{Avg}$ / $H_{best}$ |
| تصویر رمزی سیاه | 3 | 0٫00023 | 0٫00063 | 0٫00043 |
| | | 0٫000063 | 0٫000017 | 0٫00029 |
| | 4 | 0٫00056 | 0٫00021 | 0٫00036 |
| | | 0٫00002 | 0٫000001 | 0٫000005 |
| تصویر رمزی سفید | 3 | 0٫00037 | 0٫00019 | 0٫00036 |
| | | 0٫000002 | 0٫000006 | 0٫000047 |
| | 4 | 0٫00013 | 0٫00039 | 0٫000176 |
| | | 0٫000007 | 0٫000026 | 0٫0000004 |

| تصویر 256×256 | دورها | $X^2_{Avg}$ / $X^2_{best}$ | $Entropy_{Avg}$ / $Entropy_{best}$ |
|---|---|---|---|
| تصویر رمزی سیاه | 3 | – | 7٫997433 |
| | | 232٫869 | 7٫99770 |
| | 4 | 208٫382 | 7٫997446 |
| | | 231٫514 | 7٫997696 |
| تصویر رمزی سفید | 3 | 208٫109 | 7٫997523 |
| | | 224٫732 | 7٫99791 |
| | 4 | 193٫6563 | 7٫997257 |
| | | 230٫057 | 7٫997090 |

در طرح GH401 برای آزمایش همبستگی بین دو پیکسل مجاور، ما به طور تصادفی 40000 جفت پیکسل مجاور را از تصویر رمزگذاری شده به ترتیب معکوس، افقی، و جهت مورب انتخاب کرده. سپس ضریب همبستگی هر زوج را محاسبه کرده‌ایم. با مقایسه طرح GH401 با سایر طرح‌ها(جدول 13) میانگین همبستگی ضرایب برای الگوریتم رمزگذاری جدید بسیار نزدیک به صفر است. همه نتایج تایید می‌کنند که طرح GH401 ما می‌تواند همبستگی بین پیکسل های مجاور را در تصویر رمزگذاری شده حذف کند.

به‌منظور نشـان‌دادن یکنـواختی حاصـل از رمزنگـاری، طـرح پیشنهادی GH401 با سایر طرح‌ها نیز مقایسـه شـده اسـت (جـدول 12). کم‌بودن مقدار این آزمون در طرح پیشـنهادی GH401 نشـان‌دهنده توزیـع یکنواخت مقادیر خاکستری نسبت به سایر طرح‌ها است

## ۶-۴-۵- ارزیابی و تحلیل فضای کلید

اندازه فضای کلید در فرآیند رمزگذاری یک معیار مهم است و برای مقاومت در برابر حمله جستوجوی فراگیر، فضای کلید باید به‌اندازه کافی بزرگ باشد. برای هر سیستم رمزنگاری، فضای کلید تعدادی از کلیدهای ممکن قابل استفاده در فرآیند رمزگذاری را نشان می دهد. درحالت کلی اندازه فضای کلید باید بزرگتر از $2^{128}$ [31] و یا $2^{100}$ [32] باشد. در [33] عنوان شده است که در استاندارد NIST و ENISA اندازه فضای کلید باید بزرگتر از $2^{112}$ باشد.

در طرح GH401 از سه سری کلید متفاوت استفاده می‌شود. اول کلیدهای $(r, e, d, c, b, a, x_6, x_5, x_4, x_3, x_2, x_1)$ که دارای دقت تا ۱۶ رقم اعشار هستند؛ دوم کلید دوری n (تعداد تکرارهای حلقه) و سوم کلیدی که در ۵-۳ مطرح شد. اندازه فضای کلید سیستم رمزنگاری پیشنهادی $10^{96} \times 2^{128} \times n$ است. بنابراین فضای کلید بزرگ‌تر از اندازه استاندارد لازم است و می‌تواند در برابر انواع حملات جستجوی فراگیر مقاومت کند.

جدول (۸) برتری امنیت کلید طرح GH401 را نشان می‌دهد ضمن اینکه فضای کلید پیشنهادشده برای مقاومت در برابر برخی دیگر از انواع متداول حملات جستجوی فراگیر کافی است.

**جدول (۸): مقایسه کلید طرح GH401 با سایر طرح‌ها**

| Proposed | $n \times 2^{128} \times 10^{96} \approx n \times 2^{447}$ |
|---|---|
| Hosny et al.[12] (IEAHF) | $10^{96} \approx 2^{319}$ |
| Hua et al.[19] | $1.5 \times 10^{77} \approx 2^{256}$ |
| Wu et al.[34] | $10^{112} \approx 2^{372}$ |
| Dridi et al.[7] | $4.85 \times 10^{83} \approx 2^{278}$ |
| Lu et al.[33] | $2^{134}$ |
| Ullah et al.[31] | $2^{280}$ |
| Zhu et al.[24] | $2^{258}$ |
| Farah et al.[35] | $2^{186} \approx 10^{56}$ |

## ۶- زمان اجرا

زمان رمزنگاری یکی از مهم‌ترین آزمون‌ها برای ارزیابی عملکرد یک الگوریتم رمز تصویر است، علاوه بر این سرعت اجرای الگوریتم رمزنگاری یک شاخص کلیدی برای اطمینان از کاربردی بودن آن است. در طرح GH401 زمان لازم برای رمزنگاری و رمزگشایی تصاویر ۲۵۶×۲۵۶ بر مبنای میانگین ۱۰۰ تکرار با استفاده از کلیدهای متفاوت و با n=4 در جدول (۹) آورده شده است. زمان اجرا با لپ‌تاپ دارای ویندوز ۱۰، Intel(R) Core(TM)i7-4710HQ CPU@2.50GHZ، رم ۱۲GB و MATLAB R2020a انجام شده است.

**جدول (۹): زمان اجرای طرح GH401**

| تصویر ۲۵۶×۲۵۶ | میانگین زمان رمزنگاری | میانگین زمان رمزگشایی |
|---|---|---|
| Baboon 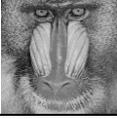 | ۰٫۲۶۳۴۰۸ | ۰٫۲۶۵۲۱۰ |
| Peppers 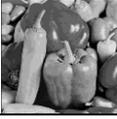 | ۰٫۲۶۴۹۴۶ | ۰٫۲۶۹۳۹۲ |
| Airplane F16 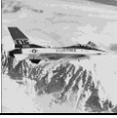 | ۰٫۲۶۳۹۷۱ | ۰٫۲۶۷۲۴۵ |
| Lena 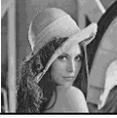 | ۰٫۲۶۵۴۶۴ | ۰٫۲۶۹۴۱۳ |
| Black 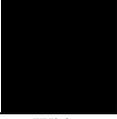 | ۰٫۲۶۲۰۲۵ | ۰٫۲۶۷۳۱۵ |
| White 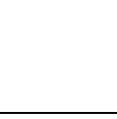 | ۰٫۲۶۵۱۵۲ | ۰٫۲۶۷۴۰۶ |

## ۷- نتیجه

در این مقاله امنیت یک طرح رمزنگاری تصویر به نام IEAHF با استفاده از نگاشت‌های آشوبی مورد مطالعه و بررسی قرار گرفت. ما نشان دادیم که این طرح دارای ضعف‌های متعدد از جمله عدم صحت رمزنگاری تصویر سیاه، رمزنگاری نامناسب تصویر سفید (نامناسب‌بودن پارامترهای آنتروپی، همبستگی، آزمون مربع خی، هیستوگرام، UACI و NPCR)، کلیدهای ضعیف، استفاده نامناسب از کلید است و تصویر رمزی حساسیت لازم نسبت به تغییرات تصویر اصلی را ندارد. علاوه بر این با بررسی خواص شبه‌تصادفی و آشوبی نگاشت‌های به‌کاررفته در طرح، نشان داده شد که الگوریتم قابل‌بهبود است و با بهبود نگاشت آشوبی ۶ بعدی، ماتریس Q-فیبوناچی، طراحی مجموعه کلیدهای جدید و استفاده از جعبه جانشانی JAYAR (که در برابر حمله DPA مقاوم است) طرح جدیدی معرفی کردیم. با مقایسه با سایر الگوریتم‌های رمزنگاری تصویر، نشان‌داده‌شده است که طرح پیشنهادی در مقایسه با طرح‌های مشابه هم از نظر امنیت و هم از نظر کارایی نتایج بهتری را تولید کرده است.

## پیوست‌ها

جدول (۱۰): مقایسه NPCR و UACI طرح GH401 با سایر طرح‌های مشابه

| Plain image | - | baboon | Peppers | Boat | AirplaneF16 | Lena | Black | White | Pass rate |
|---|---|---|---|---|---|---|---|---|---|
| **Proposed** | $NPCR_{Avg}$ | ۹۹.۶۳۸۵ | ۹۹.۶۲۹۴ | ۹۹.۶۴۳۶ | ۹۹.۶۲۰۲ | ۹۹.۶۴۵۳ | ۹۹.۶۰۹۳ | ۹۹.۶۱۴۵ | ۱۴/۱۴ |
| | $NPCR_{best}$ | ۹۹.۶۷۰۴ | ۹۹.۶۶۴۴ | ۹۹.۶۷۴۹ | ۹۹.۶۸۴۱ | ۹۹.۶۶۱۳ | ۹۹.۶۴۸۳ | ۹۹.۶۵۹۰ | |
| | $UACI_{Avg}$ | ۳۳.۴۵۶۱ | ۳۳.۴۶۲۶ | ۳۳.۴۶۶۸ | ۳۳.۴۵۷۰ | ۳۳.۴۷۴۵ | ۳۳.۴۵۹۷ | ۳۳.۴۷۶۹ | |
| Hosny et al.[12](IEAHF) | NPCR | ۹۹.۵۹۴۱ | ۹۹.۶۰۳۳ | ۹۹.۶۰۷۸ | ۹۹.۶۰۱۷ | ۹۹.۶۲۶۴ | ۰.۰۰۰۰۶ | ۷۴.۶۹۷۸ | ۱۰/۱۴ |
| | UACI | ۳۳.۴۶۱۰ | ۳۳.۴۲۷۴ | ۳۳.۴۱۸۸ | ۳۳.۵۰۵۳ | ۳۳.۴۲۲۶ | ۰.۰۰۰۰۱ | ۴۱.۵۶۰۴ | |
| Hua et al.[19] | NPCR | ۹۹.۶۳۰۷ | ۹۹.۶۲۳۱ | ۹۹.۵۶۸۲ | ۹۹.۶۲۳۱ | ۹۹.۵۸۵۰ | | | ۸/۱۰ |
| | UACI | ۳۳.۴۵۳۴ | ۳۳.۶۸۰۵ | ۳۳.۳۶۳۳ | ۳۳.۴۶۶۵ | ۳۳.۵۵۸۲ | | | |
| Wu et al.[34] | NPCR | ۹۹.۵۹۲۵ | ۹۹.۶۰۷۸ | ۹۹.۶۱۷۰ | ۹۹.۶۲۳۱ | ۹۹.۶۲۰۰ | | | ۹/۱۰ |
| | UACI | ۳۳.۳۸۲۲ | ۳۳.۴۹۵۳ | ۳۳.۶۶۰۹ | ۳۳.۶۳۵۸ | ۳۳.۴۱۶۹ | | | |
| Hanif et al.[30] | NPCR | ۹۹.۶۵۲۱ | | ۹۹.۵۹۸۷ | | ۹۹.۵۷۴۳ | | | ۳/۶ |
| | UACI | ۳۳.۱۶۲۷ | | ۳۳.۲۲۶۲ | | ۳۳.۰۵۰۹ | | | |
| Dridi et al.[7] | NPCR | | | | | | ۹۹.۶۶۷ | ۹۹.۶۰۰۸ | ۴/۴ |
| | UACI | | | | | | ۳۳.۴۸۴۰ | ۳۳.۵۱۳۹ | |
| Jun et al.[26] T=50 | NPCR | ۹۹.۶۰۸۲ | ۹۹.۶۱۳۴ | ۹۹.۶۱۵۹ | | ۹۹.۶۱۹۷ | | | ۴/۸ |
| | UACI | ۳۲.۱۳۱۶ | ۳۳.۱۹۶۴ | ۳۲.۹۶۹۳ | | ۳۳.۰۴۴۳ | | | |
| Ullah et al.[31] | NPCR | | | | | | ۹۹.۶۲ | ۹۹.۶۰ | ۴/۴ |
| | UACI | | | | | | ۳۳.۴۱ | ۳۳.۴۵ | |
| Zhu et al.[24] | NPCR | | | | | ۹۹.۵۹ | | | ۱/۲ |
| | UACI | | | | | ۳۳.۲۵ | | | |
| Farah et al.[35] | NPCR | | ۹۹.۶۲۲۳ | | ۹۹.۶۱۳۰ | ۹۹.۶۰۸۳ | | | ۳/۶ |
| | UACI | | ۳۳.۵۴۲۰ | | ۳۳.۴۷۶۶ | ۳۳.۴۳۱۲ | | | |
| Sarosh et al.[36] | NPCR | ۹۹.۶۰۴۸ | ۹۹.۶۰۶۳ | | | ۹۹.۵۸۸۰ | | | ۸/۸ |
| | UACI | ۳۳.۴۹۱۴ | ۳۳.۵۵۲۷ | | | ۳۳.۴۰۵۹ | | | |
| Pourasad et al.[37] | UACI | ۳۳.۴۱۵ | ۳۳.۶۲۱ | ۳۳.۶۷۱ | | ۳۳.۱۲۰ | | | ۲/۴ |

جدول (۱۱): مقایسه آنتروپی طرح GH401 با سایر طرح‌های مشابه

| Plain image | baboon | Peppers | Boat | AirplaneF16 | Lena | Black | White | Average |
|---|---|---|---|---|---|---|---|---|
| $Proposed_{Avg}$ | ۷.۹۹۷۵۷۶ | ۷.۹۹۷۳۹۷۳ | ۷.۹۹۷۵۹۲۵ | ۷.۹۹۷۴۷۷۵ | ۷.۹۹۷۶۶۷ | ۷.۹۹۷۴۶۱ | ۷.۹۹۷۲۵۷ | ۷.۹۹۷۴۸۴ |
| $Proposed_{best}$ | ۷.۹۹۷۹۵۲ | ۷.۹۹۷۶۸۷ | ۷.۹۹۷۸۰۰۲ | ۷.۹۹۷۷۸۷ | ۷.۹۹۷۸۱۹ | ۷.۹۹۷۶۹۶ | ۷.۹۹۷۷۰۹۰ | ۷.۹۹۷۷۹۱ |
| Hosny et al.[12](IEAHF) | ۷,۹۹۷۵ | ۷,۹۹۷۰ | ۷,۹۹۷۶ | ۷,۹۹۷۲ | ۷,۹۹۷۲ | ۰ | ۲,۷۴۷۵ | ۶.۱۰۴۸۵۷ |
| Hua et al.[19] | ۷.۹۹۷۴ | ۷.۹۹۷۱ | ۷.۹۹۷۴ | ۷.۹۹۷۲ | ۷.۹۹۷۶ | - | - | ۷.۹۹۷۳ |
| Wu et al.[34] | ۷.۹۹۷۱ | ۷.۹۹۷۴ | ۷.۹۹۷۱ | ۷.۹۹۷۱ | ۷.۹۹۷۶ | - | - | ۷.۹۹۷۲ |
| Hanif et al.[30] | ۷.۹۹۵۲ | - | ۷.۹۹۵۶ | - | ۷.۹۹۵۷ | | | ۷.۹۹۵۵ |
| Dridi et al.[7] | | | | | | ۷.۹۹۷۴ | ۷.۹۹۶۸ | ۷.۹۹۷۱ |
| Jun et al.[26] | ۷.۹۹۶۱ | ۷.۹۹۷۰ | ۷.۹۹۶۷ | - | ۷.۹۹۷۱ | - | - | ۷.۹۹۶۷۲۵ |
| Lu et al.[33] | ۷.۹۹۷۱ | - | - | - | ۷.۹۹۷۱ | | | ۷.۹۹۷۱ |
| Ullah et al.[31] | | | | | ۷.۹۹۷۴ | ۷.۹۹۶۹ | ۷.۹۹۶۹ | ۷.۹۹۷۰۷ |
| Zhu et al.[24] | ۷.۹۹۶۸ | ۷.۹۹۷۵ | | | ۷.۹۹۷۶ | ۷.۹۹۷۲ | ۷.۹۹۶۸ | ۷.۹۹۷۱۸ |
| Farah et al.[35] | | ۷.۹۹۷۲۷۵ | | ۷.۹۹۷۰۹۰ | ۷.۹۹۷۰۵۷ | | | ۷.۹۹۷۱۴۰ |
| Sarosh et al.[36] | ۷.۹۹۷۱ | ۷.۹۹۷۲ | | ۷.۹۹۷۳ | ۷.۹۹۷۰ | | | |

جدول(۱۲): مقایسه مربع خی دو طرح GH401 با سایر طرح‌های مشابه

| Plain image | baboon | Peppers | Boat | AirplaneF16 | Lena | Black | White |
|---|---|---|---|---|---|---|---|
| $Proposed_{Avg}$ | ۲۲۱.۰۹۳ | ۲۲۱.۶۸۷ | ۲۴۴.۸۷۷ | ۲۴۳.۸۲۷ | ۲۲۷.۸۴۳ | ۲۳۱.۵۱۴ | ۲۴۲.۵۷۱ |
| $Proposed_{best}$ | ۱۸۶.۹۲۱ | ۲۰۹.۳۵۱ | ۱۹۸.۸۲۸ | ۱۹۹.۸۷۵ | ۱۹۸.۳۳۵ | ۲۰۸.۱۰۹ | ۲۰۶.۵۱۵ |
| Hosny et al.[12](IEAHF) | ۲۲۴.۲۵۷ | ۲۶۸.۴۷۶ | ۲۱۹.۹۲۹ | ۲۵۳.۹۰ | ۲۶۴.۸۷۵ | ۱۶۷۱۱۶۸۰ | ۲۵۶۵۹۱۵ |
| Jun et al.[26] T=50 | ۲۷۵.۷۸۱۳ | ۲۶۱.۱۴۰ | ۲۷۸.۱۶ | | ۲۷۴.۷۱۸ | | |
| Lu et al.[33] | ۲۶۷.۸۱۲ | | | | ۲۶۲.۵ | | |
| Zhu et al.[24] | ۲۸۸.۶۶۴ | ۲۲۴.۲۳۴ | | | ۲۲۱.۱۹۵ | ۲۵۶.۴۳۰ | ۲۹۳.۰۳۹ |
| Dridi et al.[7] | | | | | | ۲۳۴.۰۳۹۱ | ۲۷۳.۶۶۴ |
| Sarosh et al.[36] | | ۲۵۶.۶۱۹ | | | ۲۳۷.۲۷۸ | | |

جدول (۱۳): مقایسه همبستگی طرح GH401 با سایر طرح‌های مشابه

| Plain image | | baboon | Peppers | Boat | AirplaneF16 | Lena | Black | White | Average |
|---|---|---|---|---|---|---|---|---|---|
| $Proposed_{best}$ | H | ۰.۰۰۰۰۳ | ۰.۰۰۰۰۱ | ۰.۰۰۰۰۴ | ۰.۰۰۰۰۸ | ۰.۰۰۰۰۹ | ۰.۰۰۰۰۵ | ۰.۰۰۰۰۴ | ۰.۰۰۰۰۴ |
| | V | ۰.۰۰۰۰۳ | ۰.۰۰۰۰۰۸ | ۰.۰۰۰۰۷ | ۰.۰۰۰۰۵ | ۰.۰۰۰۰۱ | ۰.۰۰۰۰۱ | ۰.۰۰۰۰۳ | ۰.۰۰۰۰۲ |
| | D | ۰.۰۰۰۰۲ | ۰.۰۰۰۰۰۷ | ۰.۰۰۰۰۱ | ۰.۰۰۰۰۱ | ۰.۰۰۰۰۶ | ۰.۰۰۰۰۲ | ۰.۰۰۰۰۷ | ۰.۰۰۰۰۲ |
| $Proposed_{Avg}$ | H | ۰.۰۰۰۷۶ | ۰.۰۰۰۰۳ | ۰.۰۰۰۱۱ | ۰.۰۰۰۰۱ | ۰.۰۰۰۱۱ | ۰.۰۰۰۳۶ | ۰.۰۰۰۱۷ | ۰.۰۰۰۲۲ |
| | V | ۰.۰۰۰۳۱ | ۰.۰۰۰۵۷ | ۰.۰۰۰۴۱ | ۰.۰۰۰۳۷ | ۰.۰۰۰۴۵ | ۰.۰۰۰۲۱ | ۰.۰۰۰۳۹ | ۰.۰۰۰۳۸ |
| | D | ۰.۰۰۰۷۲ | ۰.۰۰۰۳۰ | ۰.۰۰۰۱۹ | ۰.۰۰۰۸۶ | ۰.۰۰۰۰۶ | ۰.۰۰۰۵۶ | ۰.۰۰۰۱۳ | ۰.۰۰۰۴۰ |
| Hosny et al.[12](IEAHF) | H | ۰.۰۲۵۱ | ۰.۰۰۴۴ | ۰.۰۰۴۱ | ۰.۰۲۶۹ | ۰.۰۰۱۹ | ۱ | ۰.۵۰۱۹۶ | ۰.۲۲۳۴۸ |
| | V | ۰.۰۰۴۰ | ۰.۰۰۷۷ | ۰.۰۰۱۱ | ۰.۰۱۹۶ | ۰.۰۰۶۹ | ۱ | ۰.۲۴۶۵ | ۰.۱۵۱۹۹ |
| | D | ۰.۰۲۳۱ | ۰.۰۰۶۷ | ۰.۰۱۱۳ | ۰.۰۲۶۰ | ۰.۰۲۰۰ | ۱ | ۰.۲۱۴۲ | ۰.۱۵۸۳۶ |
| Hua et al.[19] | H | ۰.۰۱۳۲ | ۰.۰۱۰۰ | ۰.۰۰۲۱۴ | ۰.۰۱۵۴ | ۰.۰۲۵۰ | – | – | ۰.۰۱۷ |
| | V | ۰.۰۱۵۰ | ۰.۰۰۳۸ | ۰.۰۳۳۰ | ۰.۰۰۸۹ | ۰.۰۱۱۶ | | | ۰.۰۱۴۵ |
| | D | ۰.۰۳۰۹ | ۰.۰۳۲۱ | ۰.۰۲۱۳ | ۰.۰۰۳۱ | ۰.۰۰۲۵ | | | ۰.۰۱۷۴۲ |
| Wu et al.[34] | H | ۰.۰۰۲۹ | ۰.۰۰۰۶ | ۰.۰۰۰۳ | ۰.۰۰۲۵ | ۰.۰۰۳۲ | | | ۰.۰۰۱۹ |
| | V | ۰.۰۰۳۳ | ۰.۰۰۳۸ | ۰.۰۰۳۴ | ۰.۰۰۵۰ | ۰.۰۰۱۶ | | | ۰.۰۰۳۴ |
| | D | ۰.۰۰۶۲ | ۰.۰۰۱۰ | ۰.۰۰۱۱ | ۰.۰۰۱۲ | ۰.۰۰۲۳ | | | ۰.۰۰۲۴ |
| Hanif et al.[30] | H | | | | | ۰.۰۰۶۵ | | | ۰.۰۰۶۵ |
| | V | | | | | ۰.۰۰۱۶ | | | ۰.۰۰۱۶ |
| | D | | | | | ۰.۰۰۶۳ | | | ۰.۰۰۶۳ |
| Dridi et al.[7] | H | | | | | | ۰.۰۰۱۶۷ | ۰.۰۰۱۶۵ | ۰.۰۰۱۶۶ |
| | V | | | | | | ۰.۰۰۰۴۴ | ۰.۰۰۰۲۵ | ۰.۰۰۰۳۴ |
| | D | | | | | | ۰.۰۰۰۷۴ | ۰.۰۰۰۳۶ | ۰.۰۰۰۵۵ |
| Jun et al.[26] T=50 | H | ۰.۰۰۱۴ | ۰.۰۰۲۶ | ۰.۰۰۲۰ | | ۰.۰۰۵۹ | | | ۰.۰۰۲۹۷ |
| | V | ۰.۰۰۳۲ | ۰.۰۰۱۲ | ۰.۰۰۴۷ | | ۰.۰۰۶۴ | | | ۰.۰۰۳۸۷ |
| | D | ۰.۰۰۱۰ | ۰.۰۰۵۰ | ۰.۰۰۱۸ | | ۰.۰۰۰۳ | | | ۰.۰۰۲۰ |
| Lu et al.[33] | H | ۰.۰۰۰۷ | | | | ۰.۰۰۵۶ | | | ۰.۰۰۳۱۵ |
| | V | ۰.۰۰۰۴ | | | | ۰.۰۰۰۶ | | | ۰.۰۰۰۵ |
| | D | ۰.۰۰۱۴ | | | | ۰.۰۰۱۸ | | | ۰.۰۰۱۶ |
| Ullah et al.[31] | H | | | | | ۰.۰۰۰۱ | ۰.۰۰۲۷ | ۰.۰۰۸۰ | ۰.۰۰۳۶ |
| | V | | | | | ۰.۰۰۰۱ | ۰.۰۰۹۰ | ۰.۰۰۵۷ | ۰.۰۰۴۹ |
| | D | | | | | ۰.۰۰۰۷ | ۰.۰۰۲۰ | ۰.۰۰۳۵ | ۰.۰۰۲۰ |
| Zhu et al.[24] | H | | | | | ۰.۰۰۲۰ | | | ۰.۰۰۲۰ |
| | V | | | | | ۰.۰۰۰۳ | | | ۰.۰۰۰۳ |
| | D | | | | | ۰.۰۰۱۴ | | | ۰.۰۰۱۴ |
| Farah et al.[35] | H | | ۰.۰۱۵۹۷ | | ۰.۰۲۱۲۷ | ۰.۰۱۱۸ | | | ۰.۰۱۶۳۴ |
| | V | | ۰.۰۳۵۰۳ | | ۰.۰۱۸۶۲ | ۰.۰۱۷۳ | | | ۰.۰۲۳۶۵ |
| | D | | ۰.۰۰۹۶۷ | | ۰.۰۱۰۸۰ | ۰.۰۰۸۰ | | | ۰.۰۰۹۴۹ |
| Sarosh et al.[36] | H | ۰.۰۰۴۰ | | | | ۰.۰۰۲۱ | | | ۰.۰۰۳۰۵ |
| | V | ۰.۰۰۱۹ | | | | ۰.۰۰۹۹ | | | ۰.۰۰۵۹ |
| | D | ۰.۰۰۱۸ | | | | ۰.۰۰۱۱ | | | ۰.۰۰۱۴۵ |

# زیرنویس‌ها

[1] Topologically transitive
[2] Dense
[3] Fibonacci Q matrix
[4] Khalid M. Hosny
[5] Number of Pixels Change Rate
[6] United Average Changing Intensity
[7] Pearson's Chi-Square
[8] Goodness of Fit